# Quasi Two-dimensional Vortex Matter in ThH$_{10}$ Superhydride

Andrey V. Sadakov[1], Vladimir A. Vlasenko[1], Dmitrii V. Semenok[2,*], Di Zhou[2,*], Ivan A. Troyan[3], Alexey S. Usoltsev[1],Vladimir M. Pudalov[1,4]

[1] P. N. Lebedev Physical Institute, Russian Academy of Sciences, Moscow 119991, Russia

[2] Center for High Pressure Science & Technology Advanced Research, Bldg. #8E, ZPark, 10 Xibeiwang East Rd, Haidian District, Beijing, 100193, China

[3] Shubnikov Institute of Crystallography, Federal Scientific Research Center Crystallography and Photonics, Russian Academy of Sciences, 59 Leninsky Prospekt, Moscow 119333, Russia

[4] National Research University Higher School of Economics, Moscow, 101000, Russia

Corresponding authors: Di Zhou (di.zhou@hpstar.ac.cn), Dmitrii Semenok (dmitrii.semenok@hpstar.ac.cn).

**Abstract**

A comprehensive study of the vortex phases and vortex dynamics is presented for a recently discovered high-temperature superconductor ThH$_{10}$ with $T_C$ = 153 K at 170 GPa. The obtained results strongly suggest a quasi two-dimensional (2D) character of the vortex glass phase transition in ThH$_{10}$. The activation energy yields a logarithmic dependence $U_0 \propto \ln(H)$ on magnetic field in a low field region and a power law dependence $U_0 \sim H^{-1}$ in a high field region, signaling a crossover from 2D regime to 3D collective pinning regime, respectively. Additionally, a pinning force field dependence showcases dominance of surface-type pinning in the vicinity of $T_C$. Thermal activation energy ($U_0$), derived within thermally activated flux flow (TAFF) theory, takes very high values above $2\times10^5$ K together with the Ginzburg number $Gi$ = 0.039 – 0.085, which is lower only than those of BiSrCaCuO cuprates and 10-3-8 family of iron-based superconductor. This indicates the enormous role of thermal fluctuations in the dynamics of the vortex lattice of superhydrides, the physics of which is similar to the physics of unconventional high-temperature superconductors.

**Introduction**

The discovery of superconductivity at 203 K in the sulphur hydride H$_3$S [1] has stimulated great interests among condensed-matter physics community. Tremendous amount of work was carried out both theoretical and experimental, leading to the emergence of novel class of high-temperature superconductors (HTSC) – superhydrides [2]. New class of superconductors consists of several different structural types (Figure 1): face-centered cubic decahydrides: LaH$_{10}$ [3,4], ThH$_{10}$ [5], CeH$_{10}$ [6]; cubic body-centered hexahydrides: YH$_6$ [7,8], CaH$_6$ [9], EuH$_6$ [10]; hexagonal nonahydrides YH$_9$ [8], CeH$_9$ [6,11], ThH$_9$ [5], tetragonal LaH$_4$ [12-14], YH$_4$ [15] and A15-type polyhydrides, including La$_4$H$_{23}$ [16,17], Lu$_4$H$_{23}$ [18] and Y$_4$H$_{23}$ [19]. These structural classes of binary hydrides also cover the structures of ternary polyhydrides. Superconductivity in hydrides emerges under extreme pressures, which can be achieved statically only in diamond anvil cells (DAC). Unlike two other major families of HTSC – cuprates and iron-based superconductors (IBSC), which were thoroughly investigated in the past by all available methods, the superhydrides are the least explored class of superconductors due to physically limited access to the high-pressure zone between two diamond anvils. Amongst the most significant experimental results firmly established for superhydrides there are the observation of isotope effect, and magnetic/non-magnetic impurity scattering effect on the critical temperature ($T_c$). Still much more efforts are required to get reliable results on diamagnetic response [20,21], flux trapping [22] and resistive transitions in extreme magnetic fields [23]. Most recently, the experiments on self-field critical current allowed to tackle not only the energy gap value, but also estimate the London penetration depth in cubic $SnH_4$ [24]. However, until now there are very few works, that study the main properties of the mixed state in hydride superconductors The mixed state governs properties of any type II superconductor in magnetic fields $H > H_{C1}$. Studying the mixed state and the vortex matter in superhydrides is essential for further understanding of the physics behind the high-pressure superconductors, as well as for possible future applications of hydrides.

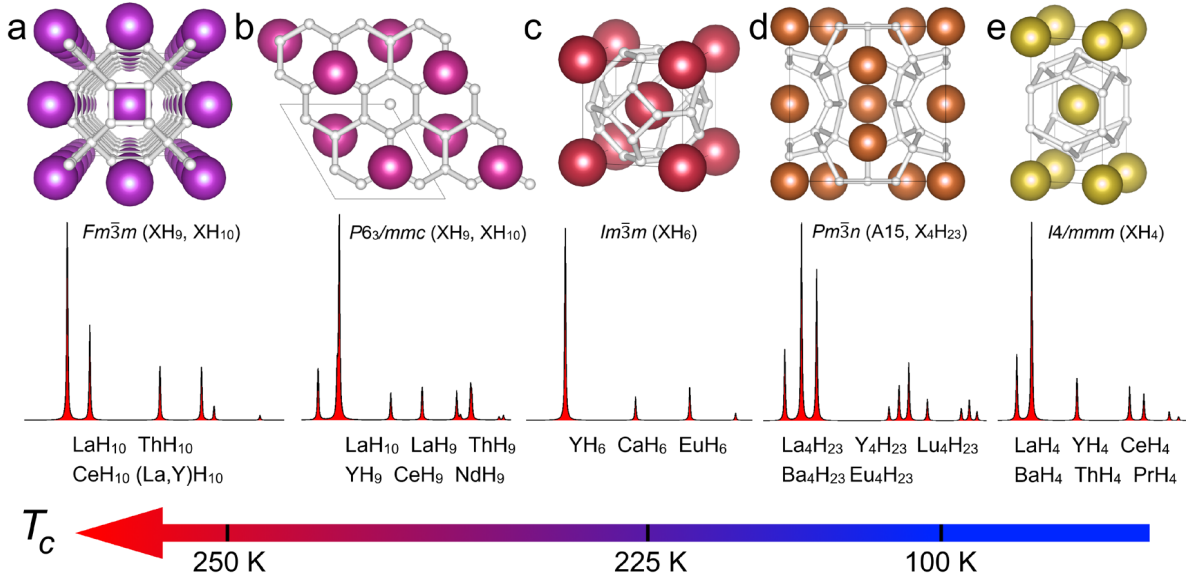

**Figure 1.** Main structural classes of superconducting metal polyhydrides with examples of their lattice structure and X-ray diffraction signatures (a) *Fcc* cubic deca- and nonahydrides with the highest known $T_C$ values ; (b) hexagonal nona- and decahydrides; (c) *bcc* cubic hexahydrides which also incorporate $H_3S$. (d) A15 class of cubic superhydrides, and (e) tetragonal superhydrides characterized by low stabilization pressure, but also by relatively low critical temperatures $T_C \approx$ 70-100 K

For cuprates and iron-based superconductors it is firmly established, that the ability to conduct current in magnetic fields without resistance in the vicinity of $T_C$ mostly depends on the thermal fluctuations. In some cases, zero resistivity in the presence of weak magnetic fields can be achieved only at very low temperatures, because of the impact of thermal fluctuations. Generally, in layered HTSC the following factors drastically increase the effect of fluctuations: high critical temperatures, short coherence length $\xi$, large penetration depth $\lambda_L$ and high anisotropy parameter $\gamma$. In superhydrides, $\xi$ is short whereas $\lambda_L$ is large, the critical temperatures are even higher, but the anisotropy ratio seem to be about unity due to highly symmetric structure similar to conventional low-temperature superconductors. In view of the above, it is therefore interesting to study, how the increase in critical temperatures up to twice as much as in cuprates and four-five times as much as in IBSC alongside with the isotropic nature of superhydrides will affect the vortex matter.

Recently, the vortex glass-liquid transitions in $Im\bar{3}m$-$YH_6$ with $T_C$ = 215 K were studied, and it was shown, that vortex motion in liquid phase is governed by thermal fluctuations, in a rather narrow (~1.5–2 K) region, but with very high activation barriers. This result seems to distinguish the yttrium superhydride $YH_6$ from other HTSC in the vortex matter physics. In order to check the generality of this result, we have chosen for studying another member of the hydrogen-rich HTSC family, thorium decahydride $ThH_{10}$ that has different critical parameters – lower critical temperature, larger coherence length and larger penetration depth.

**Experimental details and synthesis of ThH$_{10}$**

Superconducting sample of $Fm\bar{3}m$-ThH$_{10}$ was synthesized in a DAC by a pulsed Nd:YAG infrared laser (1.064 μm) heating of about 2 μm thick thorium piece loaded in ammonia borane (NH$_3$BH$_3$) medium that plays the role of a source of hydrogen. Laser heating by 4−6 pulses, with a duration of each pulse about 300 ms, was performed under pressure of about 170 GPa. The diameter of the diamond anvils was 280 μm beveled at an angle of 8.5° to a culet of 50 μm. We used a composite insulating gasket made of CaF$_2$/epoxy and a non-magnetic steel. Additional details of the synthesis, structural X-ray diffraction studies, and characterization of the sample were published in Ref. [5].

Measurements of current−voltage characteristics and electrical resistance were performed with a combination of a Keithley 6221 DC current source and a Keithley 2182a nanovoltmeter, with a standard four-probe technique in the Van der Paw scheme. The Cernox temperature sensor was fixed directly on a diamond anvil seat very close to the sample with a thermoconductive paste in order to correctly measure temperature of the sample. Experiments in magnetic fields were performed with Cryogenic CFMS-16 system.

**Results and discussion**

*Vortex glass and linear resistivity in ThH$_{10}$*

According to the vortex glass theory proposed by Fisher et al. [25,26] for type II superconductors the analysis of the linear current-voltage characteristics ($V \propto I$) in a low current mode and corresponding temperature dependencies of the (linear) resistivity in magnetic fields may reveal the presence of a vortex glass. In the vicinity of the vortex-glass transition temperature $T_g$, the vortex matter in type II superconductors with quenched disorder undergoes a second order phase transition. This phase transition is characterized by the vortex-correlation length $\xi_g$ and vortex-relaxation time $\tau_g$ of the fluctuations of the glassy order parameter, and they both diverge at $T_g$. The correlation length diverges as $\xi_g \propto |1 - T/T_g|^{-\nu}$, and the relaxation time diverges as $\tau_g \propto (\xi_g)^z \propto |1 - T/T_g|^{-z\nu}$, where $\nu$ is the static exponent, and $z$ – the dynamic exponent. Then, linear resistivity vanishes following the

$$\rho_{lin}(T) \propto |T - T_g|^s, \qquad (1)$$

where s = $\nu \times (z + 2 - d)$ and $d$ is the sample dimensionality [25,26]. On the other hand, current-voltage characteristics follow a scaling law, in which $I$-$V$ isotherms below $T_g$ and above $T_g$ collapse into two distinct scaling curves in coordinates $V_{scale}$ versus $I_{scale}$:

$$V_{scale} = (V/I) \times |1 - T/T_g|^{-\nu(z+2-d)} \, ; \, I_{scale} = (I/T) \times |1 - T/T_g|^{-\nu(d-1)}. \qquad (2)$$

*Linear resistivity measurements.*

We start with investigation of the linear resistivity. Figure 2a shows temperature dependence of resistivity of ThH$_{10}$ sample, measured across the superconducting transition, at 170 ± 5 GPa in various magnetic fields up to 7 Tesla. One can see that in zero magnetic field the sample experiences a superconducting transition at about $T_C$ = 153 K ($\rho_{50}$ criteria): below $T_C$ the resistivity drops quickly to zero. The transition shifts to lower temperatures and slightly broadens as external magnetic field increases. The studied ThH$_{10}$ sample had a shape of a disk with ~2 μm thickness and ~ 20-25 μm diameter. In our measurements we used a current of ~ 100 μA, and the resistivity was estimated according to the van der Pauw formula $\rho = R\pi t/ln(2)$, where $t$ – is the sample thickness, $R$ – is the electrical resistance of sample.

In order to investigate the vortex glass transition in ThH$_{10}$, we first analyzed the low-dissipation (linear) part of the resistive transitions. In Figure 2b, upper panel, we show ln($\rho$) vs $T$ curves for two magnetic field values. On the lower panel of this figure we show the inverse derivative (dln($\rho$)/d$T$)$^{-1}$ vs $T$, recalculated for the same curves. One can see a linear region, showcasing the power law dependence reminiscent of eq. (1) with $T_g$ defined as an intercept with $T$-axis, and the slope of the line being 1/$s$. The temperature, where the inverse derivative deviates from linear behavior is marked as $T^*$, and it signifies an upper bound of the critical region, where the second order vortex phase transition takes place.

A complete set of inverse derivatives for all magnetic fields is shown in Figure 2c. The values of the critical exponent for different fields vary approximately from 1.4 to 2.0 (see inset of Figure 2c). These values are comparable to those, obtained for cuprate HTSC, for example for heavily irradiated sample of YBCO in Ref. [27], where authors also obtained critical exponent ranging from $s$ =1.2 to 3 (for similar magnetic fields). According to the modified vortex-glass model, suggested by Andersson et al. [28,29], normalized linear resistivity in various magnetic fields in the critical region should scale into one universal curve in coordinates $\rho/\rho_n$ vs $T_{scale}$ = [$T(T_C - T_g)/T_g(T_C - T) - 1$], where $\rho_n$ is a resistivity in the normal state near superconducting transition. As can be seen in Figure 2d, the resistivity transitions in fields between 0.5 and 7 T scale into one curve with the critical exponent $s$ = 1.3. Excellent agreement with the Andersson model for ThH$_{10}$ indicates the applicability of vortex dynamics concepts to the hydride superconductors ThH$_{10}$ at ultrahigh pressures.

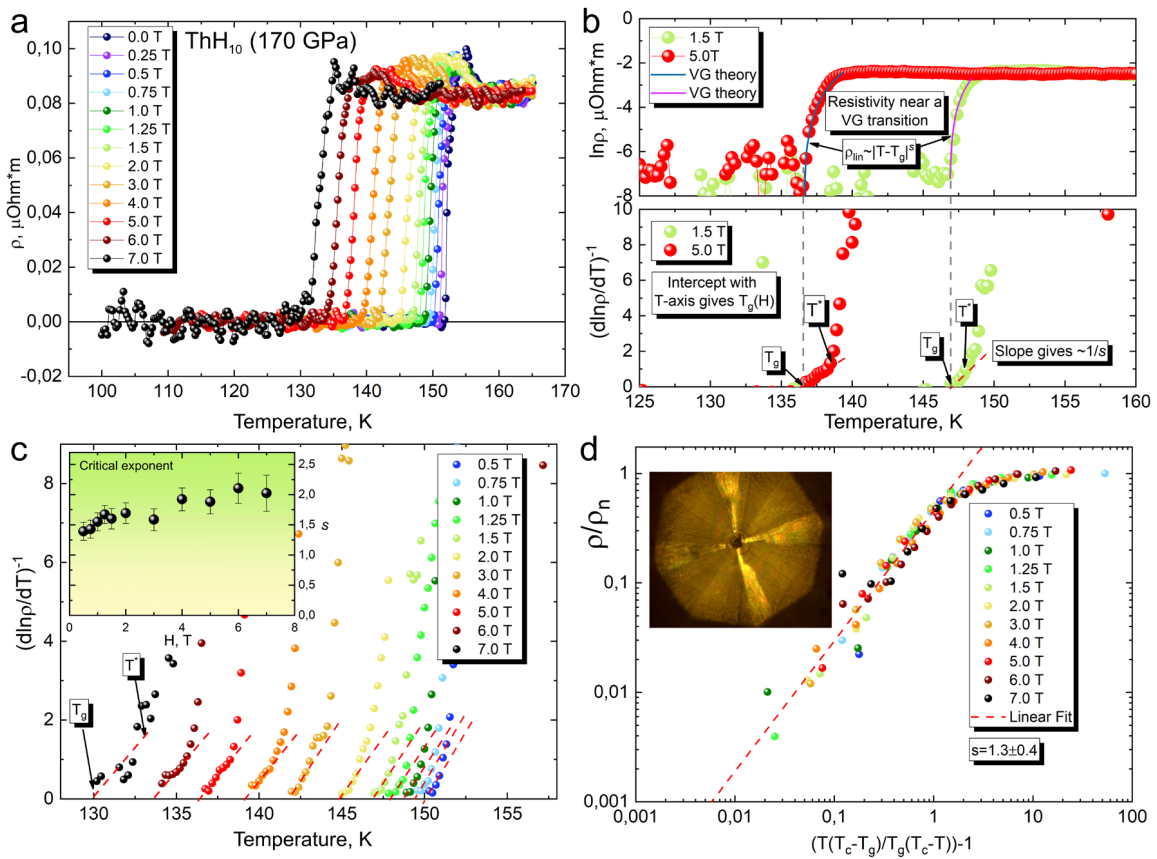

**Figure 2.** Resistive transitions in thorium hydride ThH$_{10}$ at 170 GPa. (a) In magnetic fields from 0 to 7 Tesla (linear scale). (b) Interpolation of temperature dependence of resistivity and its temperature derivative (logarithmic scale) using vortex glass (VG) model $\rho_{lin}(T) \propto |T - T_g|^s$ at 1.5 and 5 Tesla. Corresponding extrapolation of the model to the intersection with the temperature axis (T-axis) gives us the melting temperature ($T_g$) of the vortex glass in ThH$_{10}$. (c) Full set of the inverse temperature derivatives of ln($\rho$) for applied magnetic fields from 0.5 to 7 Tesla. Red dotted lines correspond to linear interpolation near $T_g$. (d)

Universal scaling of relative resistivity versus rational function of $T_C$, $T_g$ and $T$ in magnetic fields of 0.5-7T. Inset: photo of the $ThH_{10}$ sample with electrical leads.

*Temperature and field dependent I-V characteristics*

It is worth noting, that even though power law behavior and scaling law of linear resistivity give significant evidence for the existence of the phase transition from vortex liquid to glass state, the most crucial and well-established proof for VG model is the scaling of current-voltage (*I-V*) characteristics. To explore this, we performed two sets of *I-V* experiments. In the first, a more traditional set, we stabilized magnetic field, and then measured *I-V* curves at different temperatures (*T*-set). In the second experiment we stabilized temperature and measured *I-V* curves in different magnetic fields (*H*-set).

The inset of the Figure 3a depicts a set of current-voltage characteristics of $ThH_{10}$ in a double logarithmic scale for a magnetic field $H$ = 2 Tesla in the temperature range from 139 K to 146 K. $T_g$ can be approximately estimated from these $\log(I)$-$\log(V)$ isotherms as the temperature at which the curvature changes from upturn to downturn with decreasing temperature. The "$T = T_g$" curve is represented by dashed line. Then, we applied universal scaling for all *I-V* curves in coordinates of eq. (2). As was found in the previous section, the linear resistivity data give rather small values of critical exponent $s$ = 1.3±0.4 (from the universal scaling, see Figure 2d) and 1.4 – 2.0 (from the inverse derivative data, Figure 2c). Such small values are usually inherent in a two-dimensional (2D) VG case [26], leading us to start scaling with parameter $d$ = 2 in eq. (2). As one can see (main panel of Figure 3a), all ten *I-V* curves for $d$ = 2 collapse into two distinct branches for $T < T_g$ and $T > T_g$. The dashed line corresponds to a power law dependence $V \propto I^{(z+1)/(d-1)}$, for the $T = T_g$ case. From this scaling we obtain the static and dynamic exponents: $v$ = 1.90 and $z$ = 0.41. The vortex glass transition temperature for $H$ = 2 T was found to be 144.7 K, which is very close to $T_g$ obtained from resistivity data.

The second set of current-voltage characteristics, which was measured at fixed temperature and in different magnetic fields (*H*-set) is shown in Figure 3b (inset) in double-logarithmic scale. Here, we can estimate the vortex glass transition field $H_g$, at which *I-V* dependences change their curvature (marked by dashed line). The main panel of Figure 3b shows that the *I-V* curves for $d$ = 2 are collapsing into two distinct brunches for $H < H_g$ and $H > H_g$. The dashed line corresponds to a power law dependence $V \propto I^{(z+1)/(d-1)}$, for the $H = H_g$ case. Resulting scaling parameters are: $v$ = 2.15 and $z$ = 0.53. The vortex glass transition field for $T$ = 148.4 K was found to be $H_g$=1.15 T. The critical exponent $s = v(z + 2 – d)$ for two sets of *I-V* curves are 0.78 and 1.14, which is rather close to the value of $s$ = 1.3 obtained from linear resistivity scaling (Figure 2d) and $s$ = 1.4 – 1.5 obtained from inverse derivatives in low fields (Figure 2c).

The scaling parameter $v$ (static exponent), determined from this analysis, ranges from 1.90 to 2.15 that is consistent with the bulk vortex glass models ($v$ varying from 1 to 2) [26]. At the same time, the dynamic exponent $z$ appears to lie between 0.41 and 0.53, which is much lower than commonly reported values, falling in the range from 4 to 7. We note, however, that similar small values of $z$ are typical for several cases, reported in quasi-2D systems: FeSe thin film [30] ($z$ varying from 0.43 to 0.56), $Bi_2Sr_2Ca_2Cu_3O_{10+\delta}$ sub micro bridges [31] ($z$ varying from 0.38 to 0.39), $YBa_2Cu_3O_x$ thin films [32], and bulk $FeSe_{1-x}S_x$ layered system [33] ($z$ = 0.82 for 6T).

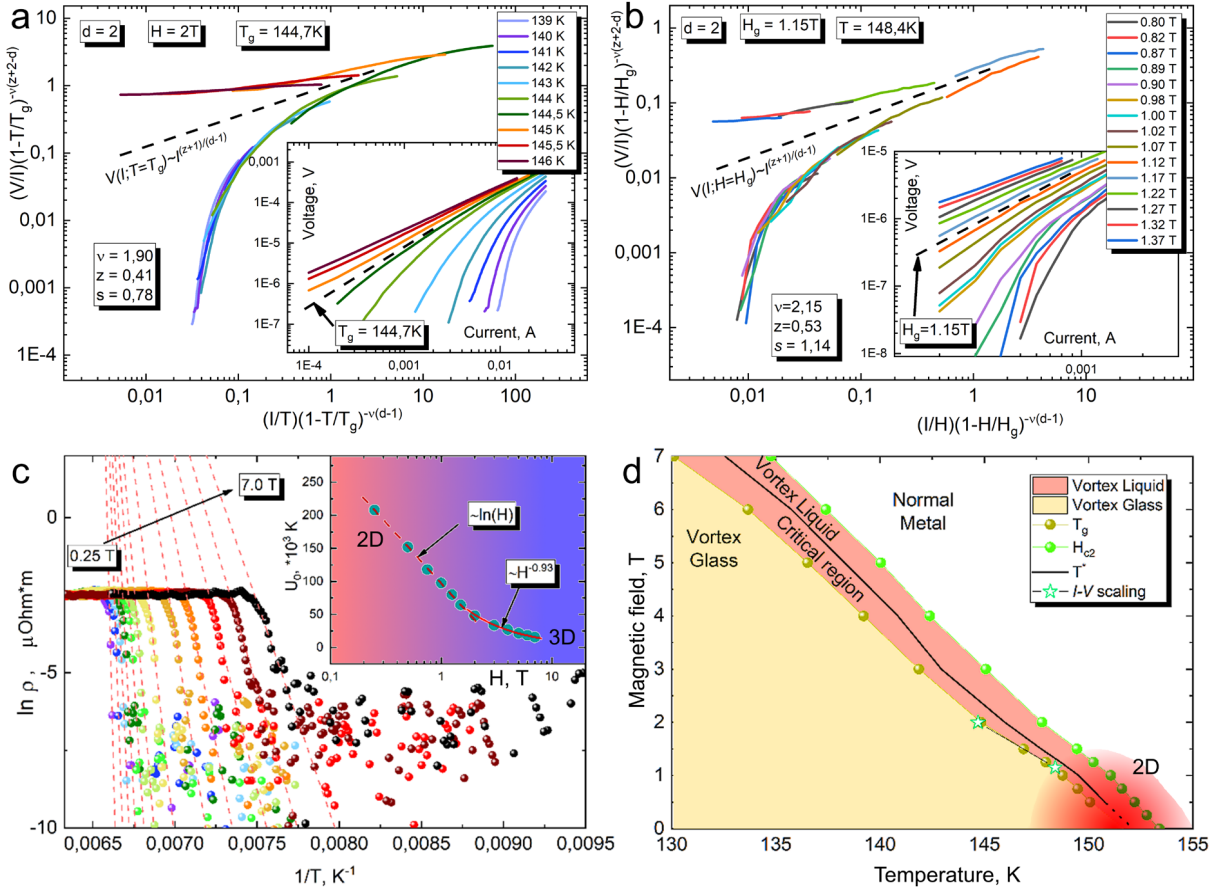

**Figure 3.** Quasi two-dimensional dynamics of the vortex lattice in $ThH_{10}$. (a) Current-voltage characteristics of $ThH_{10}$ in the double logarithmic scale of eq. (2) for a magnetic field $H = 2$ Tesla in the temperature range from 139 K to 146 K (*T*-set). Inset: *I-V* characteristics in a double logarithmic scale. (b) Current-voltage characteristics of $ThH_{10}$ in the double logarithmic scale at 148.4 K in the range of magnetic fields from 0.8 T to 1.37 T (*H*-set). The scaling for *H*-set of *I-V* curves is performed in the same manner as the one for *T*-set with isotherms, but in this case, the scaling coordinates [eq. (2)] will have temperature dependent term $T/T_g$ substituted by the field-dependent term $H/H_g$ in accordance with similar vortex glass investigations [34,35]. (c) Arrhenius plot $\ln(\rho)$ versus $1/T$, where $\rho$ – is resistivity of the sample with assumed thickness of 2 μm. The dashed red lines are fitting results from the Arrhenius relation. Inset: field dependence of apparent activation energy $U_0 = -d(\ln \rho)/d(1/T)$ (y-axis) and magnetic field (x-axis) in semi logarithmic scale. (d) Vortex phase diagram of $ThH_{10}$ at 170 GPa.

In order to double check the above quasi-2D scaling we applied critical scaling procedure with $d = 3$. The results of the scaling are as follows: $v = 0.88$ and $z = 1.74$ (see Supplementary Materials, Figure S1). As one can see, the scaling results for $d = 3$ are much less consistent with the linear resistivity data ($s = v(z + 2 - d) = 0.6$ for *T*-set) and are inconsistent with generally expected values of the critical exponents for 3D case. Usually, in three-dimensional case, $v$ varies from 1 to 2, $z$ belongs to the 4 – 7 interval, and $s$ lies between 2.7 and 9 [26].

Summarizing the results of this section, we can state that according to the linear resistivity scaling, inverse derivative data and universal critical scaling for two sets of *I-V* curves, the vortex glass state emerges in the $ThH_{10}$ superconductor as a 2D-state, at least in fields up to 2T. The obtained values of the critical exponents (*s*, *v* and *z*) are similar to those in several preceding experiments where quasi-2D vortex glass was studied in highly anisotropic unconventional HTSC [31], and also in systems with rather low anisotropy (like FeSe and $FeSe_{1-x}S_x$ [31]). The low dimensionality of vortex system occurs in superconductor when vortex correlation length becomes comparable with either one of sample dimensions, distance between weakly coupled superconductor planes (like in most cuprates and some iron-based HTSC) or distance between planar defects, such as grain boundaries, with absent or suppressed superconductivity. We believe, the latter being the likely scenario for ThH$_{10}$. Grain boundaries, play a role of normal core surface-like defects and effectively break the 3D correlation between vortices. However, the existence of the planar defects is very difficult to establish without scanning technique, like–TEM [36]. However, we performed a detailed investigation of pinning force and critical current field dependences, which can give us information of the type of defects and geometry of pinning centers.

According to flux-pinning theory, developed by Dew-Hughes [37], the pinning force $F_p = I_c*H$ has a strictly defined functional dependence for each of the individual types of defect geometry: volume, surface or point-like. The details of the pinning force investigation are presented in Supplementary Materials, Figure S2. The results show, that in the vicinity of $T_C$ pinning force follows the functional dependence $F_p \propto h^{0.6}(1 - h)^{2.1}$, with maximum at $h_{max} = 0.22$ ($h = H/H_{irr}$ is the reduced field, and $H_{irr}$ - the irreversibility field). This dependence is in very good agreement with the one $F_p \propto h^{0.5}(1 - h)^{2.0}$ (where $h_{max} = 0.2$) theoretically predicted for normal core surface like pinning centers. Thus, our analysis of the pinning force indicates the presence of surface defects in the sample, which lead to the 2D character of the vortex matter in ThH$_{10}$.

*Vortex liquid and TAFF theory*

Above the melting line, the vortex glass undergoes a phase transition into the vortex liquid state. Here the dynamics of the system is regulated by the competing contributions of vortex motion, guided by thermal energy $k_BT$ and the effective pinning energy $U(T, H)$. If the pinning energy is smaller than the temperature, they can be neglected and then the system is in unpinned vortex liquid state. However, if $U(T, H)$ is higher than the temperature, then energy barriers play vital role in vortex motion via thermally activated mechanism and the system is in the TAFF state. According to the thermally-activated flux flow (TAFF) theory [26,38], if the applied current is small enough then a temperature dependence of resistivity is defined as

$$\rho(T,H) = \rho_0 \exp(-U(T, H)/T), \qquad (3)$$

where $U(T, H)$ is the activation energy, and prefactor $\rho_0$ is in general case dependent on both temperature and magnetic field. However, for the most of experimental studies in iron-based superconductors, cuprates, and conventional superconductors it is assumed, that $\rho_0$ is temperature independent, and the activation energy is defined as $U(T, H) = U_0(H) \times (1 - T/T_C)$. Here $U_0$ is the apparent activation energy. With these two assumptions, natural logarithm of the linear electrical resistivity (eq. (3)) is expressed as:

$$\ln \rho(T,H) = \ln \rho_0 - U_0(H)/T + U_0(H)/T_C. \qquad (4)$$

Equation (4) is known as the Arrhenius relation, and the plot in $\ln(\rho)$ vs $1/T$ coordinates will exhibit a linear behavior in the TAFF regime [33,39]. In Figure 3c, we present $\ln(\rho)$ vs $1/T$ plots for different magnetic fields from 0.25 T to 7 T, in order to find the temperature region that satisfy the Arrhenius relationship (eq. (4)). As figure 3 shows, for each magnetic field, there is a distinct linear region marked by red dashed lines, following TAFF regime, and its slope gives us the activation energy $U_0(H)$. The magnetic field dependence $U_0(H)$ is presented in the inset of Figure 3c. As one can see, the flux pinning activation energy shows two distinct regions in semi-logarithmic scale. In one region, from 0.25 T up to 1.5 T, it exhibits a logarithmic behavior $U_0(H) \propto - \ln(H)$, and at higher fields we see a crossover to $U_0(H) \propto 1/H$ behavior, thus suggesting different mechanism for activated vortex motion in these field regions. We estimated the flux flow barriers $U_0(H) = 1.5-20 \times 10^4$ K to be very high (for H=2T), even higher than in YH$_6$ [40].

Usually the $U_0(H) \propto -\ln(H)$ behavior is observed in many 2D [41-43] and quasi 2D systems [44,45], in agreement with the results of the previous sections. There are several possible mechanisms which can be responsible for these particular field dependences of the activation energy. First mechanism that leads to logarithmic behavior of pinning energy is the motion of thermally activated vortex-antivortex pairs, observed, for example, in highly anisotropic Bi-2212 cuprates [46], where interplanar coupling is particularly weak. It was proposed by Jensen et al [47], that this type of vortex movement is associated with activation energy

$$U_{V-V^a} = \frac{\phi_0^2 d}{4\pi \lambda_L(T)^2} \ln\left(\frac{a_0}{\xi(T)}\right), \quad (5)$$

where $\lambda_L$ and $\xi$ are the London penetration depth and coherence length, respectively, $\phi_0$ is a flux quantum, $d$ is the thickness of the superconductor, and $a_0 = (\phi_0/H)^{0.5}$ is the flux-line spacing. Equation (5) describes logarithmic decrease of the activation energy with magnetic field. However, we believe it is highly unlikely, that vortex-antivortex case, which was only observed in very specific quasi-2D systems, is applicable for polycrystalline hydride system like our $ThH_{10}$ sample.

The second mechanism that leads to logarithmic behavior of pinning energy is the nucleation of edge dislocation pairs in the vortex system. Following Ref. [48], the energy, required to unbind a small dislocation pair is finite and leads to an activation energy given by

$$U_{edge} = \frac{\phi_0^2 d}{16\pi \lambda_L(T)^2} \ln\left(\frac{a_0}{\xi(T)}\right), \quad (6)$$

which is very similar to eq.(5) and leads to similar $U_0(H) \propto -\ln(H)$ dependence.

Next possible mechanism of dissipation is the vortex cutting and reconnection in the entangled vortex liquid, proposed in ref. [49]. Entangled vortices can cut each other, if the distance between them decreases below the average flux-line spacing $a_0$ and becomes comparable to the vortex core diameter $\xi$. The energy required for cutting (neglecting the energy due to the vortex elongation) is

$$U_{cut} = \frac{\phi_0^2 d}{32\pi^2 \lambda_L(T)^2} \ln\left(\frac{a_0}{\xi(T)}\right), \quad (7)$$

As one can see from eqs. 5-7 the pre-factors before logarithm (with slightly different numerical multipliers) are dependent on the superconductor thickness, $d$, and London penetration depth. Following the steps in Ref. [45], we now estimate the correlation length $\xi_g$ for our vortex system, using the experimental slope of activation energy $U \propto \ln(H)$ dependence, and London penetration depth obtained from self-field critical current experiments in accordance with refs. [50,51] and also ref. [24] (for details, see Supplementary Materials, Figure S3). Resulting values of the correlation length $\xi_g$ are: 6.5 μm and 13 μm for eqs. (6-7), respectively. Both correlation length estimations exceed our sample thickness ($d \sim 2$ μm), validating the assumption of quasi-2D behavior of vortex liquid in $ThH_{10}$ in weak magnetic fields. However, within our investigations, we are unable to exclude any of the two possible mechanisms for thermally activated dissipation.

The power law behavior of activation energy in higher fields $U \propto 1/H$ (as shown on the lower inset in Figure 3c) starts at approximately 1.5 Tesla and goes up to 7 Tesla. The $H^\alpha$-dependence with $\alpha = -1$ behavior is rather typical for high-$T_C$ cuprate superconductors [45,52], iron based superconductors [53] and boride superconductors [54]. Such power law is associated with three-dimensional collective pinning regime [53,55,56]. In this regime, when magnetic fields are high enough, so that the intervortex interaction becomes significant, and $U_0$ is controlled by collective pinning of the vortex bundles confined by a field-dependent correlation volume [57,58]. Thus, our results depict a crossover in the vortex activation energy from two-dimensional regime, controlled by either vortex edge dislocations or vortex cutting, to three-dimensional regime, controlled by collective pinning of the vortex bundles.

Finally, based on the values of $H_{C2}$(T), $T_g(H)$, $T^*(H)$, a detailed $H$-$T$ vortex phase diagram is constructed in Figure 3d, reflecting all the experimental data and theoretical investigations of linear resistivity, universal $I$-$V$ scaling and thermally activated vortex motion. We were able to distinguish four different regions in the mixed state of $Fm\bar{3}m$-ThH$_{10}$ superconductor: (I) Vortex glass region, displaced below the melting line $T_g(H)$; (II) Critical region between $T_g(H)$ and $T^*(H)$ lines, where a gradual phase transition takes place; (III) quasi 2D TAFF region above the $T^*(H)$ line in fields $H < 1.5$ T; (IV) 3D collective pinning region. Using $T_C = 153$ K, $H_{C2}(0) = 40$ T and ξ = 2.87 nm, we estimated the Ginzburg number for ThH$_{10}$ as $Gi = 0.039 - 0.085$. Such large values are typical for cuprates [26] and iron-containing pnictides [59], indicating a direct analogy in the dynamics of vortex matter between unconventional HTSC superconductors and thorium ThH$_{10}$ and yttrium YH$_6$ [40] superhydrides.

## Conclusions

To summarize, we have studied the complex vortex state in a ThH$_{10}$ high temperature superconductor under 1.7 Mbar pressure. Our results suggest two-dimensional nature of the vortex glass state, reminiscent of copper-based superconductors and some other quasi-2D systems. Within the framework of TAFF theory we established that vortex motion in liquid state is governed by strong thermal fluctuations. Thermal activation energy ($U_0$) and the Ginzburg number are very high in ThH$_{10}$: $U_0$ exceeds $2\times10^5$ K and $Gi = 0.039 - 0.085$, respectively. In terms of these parameters, the thorium hydride approaches the records of BiSrCaCuO system and Ca$_{10}$(Pt$_3$As$_8$) [(Fe$_{1-x}$Pt$_x$)$_2$As$_2$]$_5$ [60]. Magnetic field dependence of activation energy exhibits a crossover at field $H_{cr} = 1.5$ T from a quasi-2D logarithmic behavior $U_0 \propto \ln(H)$ to 3D power law behavior $U_0 \propto 1/H$, apposite for collective pinning regime. The analysis of low-dissipation region of linear resistivity in combination with universal scaling of current-voltage characteristics allows us to assert the existence of the vortex glass state, its melting line $T_g(H)$ and critical region in accordance with the VG theory.

## Acknowledgments

A.V.S. thanks the financial support of RSF Grant 22-22-00570. D.S. and D.Z. thank National Natural Science Foundation of China (NSFC, grant No. 1231101238) and Beijing Natural Science Foundation (grant No. IS23017) for support of this research. The work by A.S.U and V.M.P. was within the State Assignment of the LPI.

SUPPLEMENTARY MATERIALS

# Two-dimensional Vortex Matter in ThH$_{10}$ Superhydride

Andrey V. Sadakov[1], Vladimir A. Vlasenko[1], Dmitrii V. Semenok[2,*], Di Zhou[2,*], Ivan A. Troyan[3], Alexey S. Usoltsev[1], Vladimir M. Pudalov[1,4]

[1] P. N. Lebedev Physical Institute, Russian Academy of Sciences, Moscow 119991, Russia

[2] Center for High Pressure Science & Technology Advanced Research, Bldg. #8E, ZPark, 10 Xibeiwang East Rd, Haidian District, Beijing, 100193, China

[3] Shubnikov Institute of Crystallography, Federal Scientific Research Center Crystallography and Photonics, Russian Academy of Sciences, 59 Leninsky Prospekt, Moscow 119333, Russia

[4] National Research University Higher School of Economics, Moscow, 101000, Russia

Corresponding authors: Di Zhou (di.zhou@hpstar.ac.cn), Dmitrii Semenok (dmitrii.semenok@hpstar.ac.cn).

## Content

1. Universal scaling of the *I-V* characteristics

According to vortex glass theory [1,2], at the vicinity of the vortex-glass temperature $T_g$, current-voltage characteristics, follow a scaling law, in which *I-V* isotherms below $T_g$ and above $T_g$ collapse into two distinct scaling curves in coordinates $V_{scale}$ vs $I_{scale}$: $V_{scale} = (V/I)|1-T/T_g|^{-\nu(z+2-d)}$; $I_{scale} = (I/T)|1-T/T_g|^{-\nu(d-1)}$.

In Figure S3 we show the scaling for d = 3, the critical exponents for these scaling are inconsistent with the expected from VG theory ($\nu = 1 \div 2$; $z = 4 \div 7$; $s = 3 \div 9$).

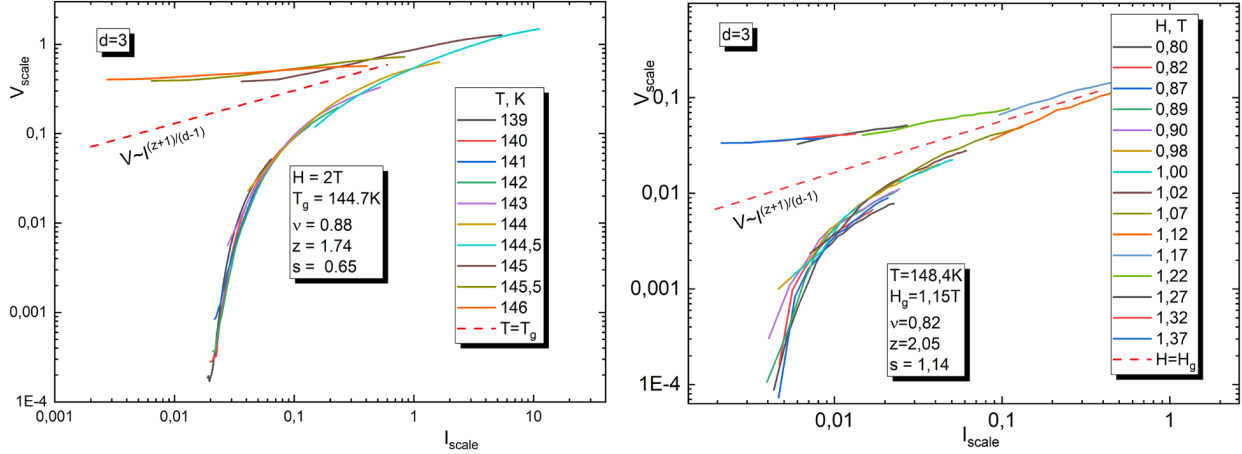

**Figure S1.** Three-dimensional vortex glass scaling of *I-V* curves for $ThH_{10}$ superconductor. Left panel: scaled *I-V* curves, measured at constant magnetic field of *H* = 2 Tesla with different temperatures (*T*-set). Right panel: scaled *I-V* curves, measured at constant temperature of *T* = 148.4 K with different magnetic fields (*H*-set).

## 2. Pinning force in ThH$_{10}$

A powerful tool to analyze the origin of pinning centers governing the critical current density dependence of type-II superconductors on applied magnetic field is to plot normalized pinning force $f_p = F_p/F_p^{max}$ as a function of the reduced field $h = H/H_{irr}$, where $H_{irr}$ - is the fields at which $F_p$ and $J_C$ fall to zero [3].

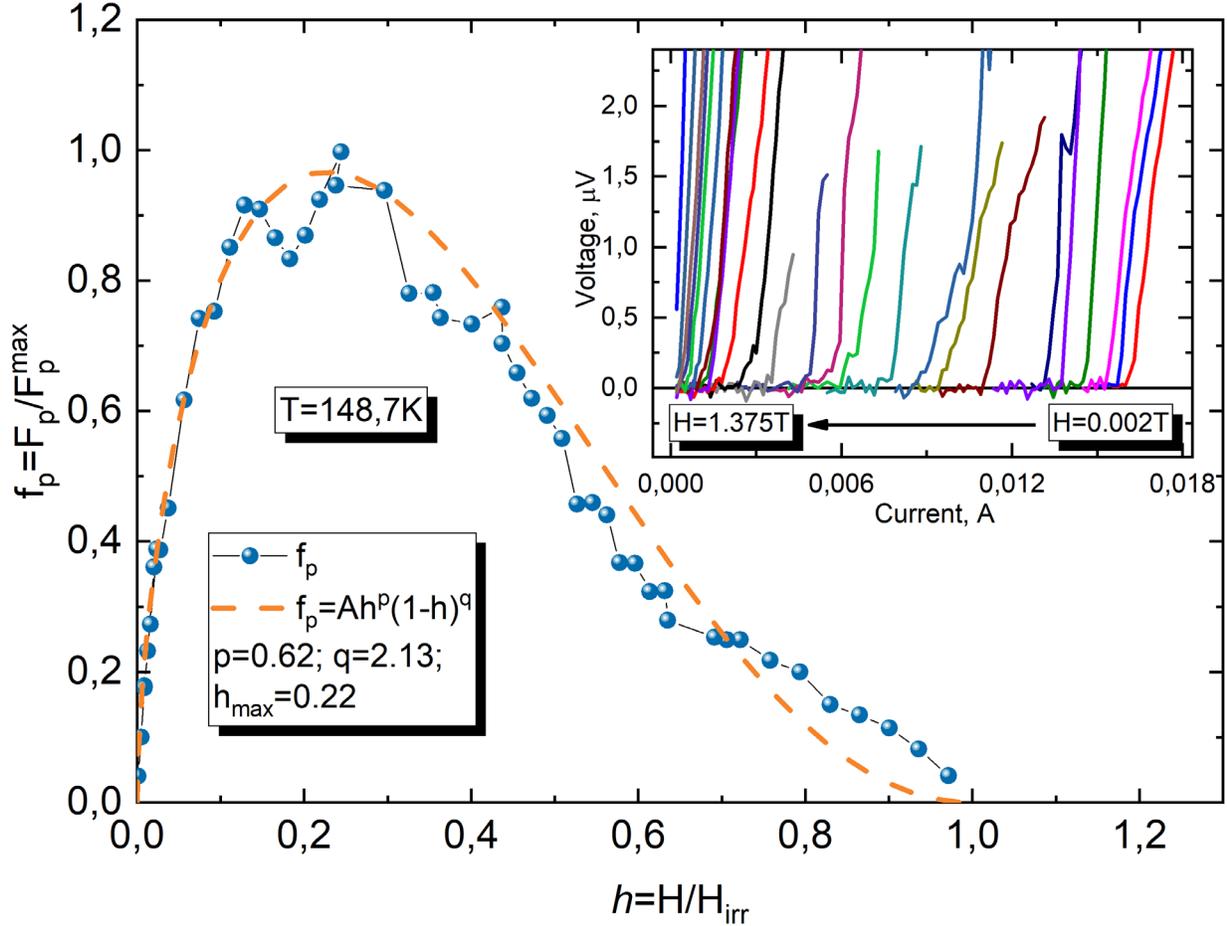

**Figure S2.** Main panel: the plot of the normalized pinning force $f_p = F_p/F_p^{max}$ versus the reduced field $h = H/H_{irr}$ are shown for temperature $T$ = 148.7 K. Experimental data is fitted by the Dew-Hughes model for surface type normal pinning centers $f_p \propto h^p(1-h)^q$. For this model parameters are $p$ = 0.5, $q$ = 2, $h_{max}$= 0.2, which is very close to our fit (shown in the figure). Inset: the *I-V* characteristics in magnetic fields up to 1.375 Tesla.

## 3. London penetration depth estimations

As was originally proposed for thin films [4] and then developed for bulk samples [5] self-field critical current is a fundamental property of a superconductor and its temperature dependence carries the information about London penetration depth and superconducting energy gap. According to [5] for type II superconductor of a rectangular cross-section shape self-field critical current is described as

$$J_c(T) = \frac{\hbar}{4e\mu_0 \lambda_L^3(T)} (\ln(\kappa) + 0.5)$$
$$\cdot \left( \frac{\lambda_L(T)}{a} \tanh\left(\frac{a}{\lambda_L(T)}\right) + \frac{\lambda_L(T)}{b} \tanh\left(\frac{b}{\lambda_L(T)}\right) \right),$$

$$\frac{\lambda_L(T)}{\lambda_L(0)} = \sqrt{1 - \frac{1}{2k_BT} \int_0^\infty \cosh^{-2}\left(\frac{\sqrt{\varepsilon^2 + \Delta^2(T)}}{2k_BT}\right)},$$

$$\Delta(T) = \Delta(0) \cdot \tanh\left( \frac{\pi k_B T}{\Delta(0)} \sqrt{\eta \left(\frac{\Delta C}{C}\right)\left(\frac{T_C}{T} - 1\right)} \right),$$

(S1)

where $2a$ – is the width of sample, $2b$ – is the thickness of sample, $\mu_0$ is the permeability of free space, $e$ – is the electron charge, $\kappa = \lambda_L/\xi$ is the Ginsburg-Landau parameter, $\Delta(T)$ – the superconducting gap, $\eta = 2/3$ for $s$-wave superconductivity, and $\Delta C/C$ – is the specific heat capacity jump at the superconducting transition. In these equations, parameters $b$, $\Delta(0)$, $\lambda_L(0)$ and $\Delta C/C$ are refined parameters. Finally, we get an approximation formula with only four fitting parameters: $T_C$, $\Delta(0)$, $\lambda_L(0)$, $\Delta C/C$. The main error for the $I$-$V$ determination of the London penetration depth comes from the inaccuracy in the absolute values of critical current density, which in its turn mostly depends on the errors in sample and contact geometry. We assume the sample thickness to be ~2 μm, and length of the contact 4-5 μm.

In Figure S3a we showed the self-field critical behavior (main panel) for cubic $ThH_{10}$ and corresponding current-voltage characteristics (inset). Critical current density is shown in in Figure S3b (purple symbols). Purple section for the critical current density theoretical curves spreads between lines, corresponding to 2×4 μm cross section (upper line) and 2×5 μm cross section (lower line). From these $J_C(T)$ curves the $\lambda_L(T)$ dependences were calculated. The obtained $\lambda_L(0)$ values are 458 nm (for 2×4 μm sample size) and 554 nm (for 2×5 μm sample size). Therefore, a roughly estimated London penetration depth is $\lambda_L(0) = 500 \pm 50$ nm. This is approximately 10-12 times higher than the results of micro magnetometry in similar superconducting materials ($H_3S$ and $LaH_{10}$) [6]. Recently, doubts have been expressed about the accuracy of determining the penetration depth and amplitude of the magnetic moment by this method [7-9]. However, it should be taken into account that the calculations in Figure S3 were made on the basis of a very small segment of the $I_C(T)$ curve and must be refined in the future.

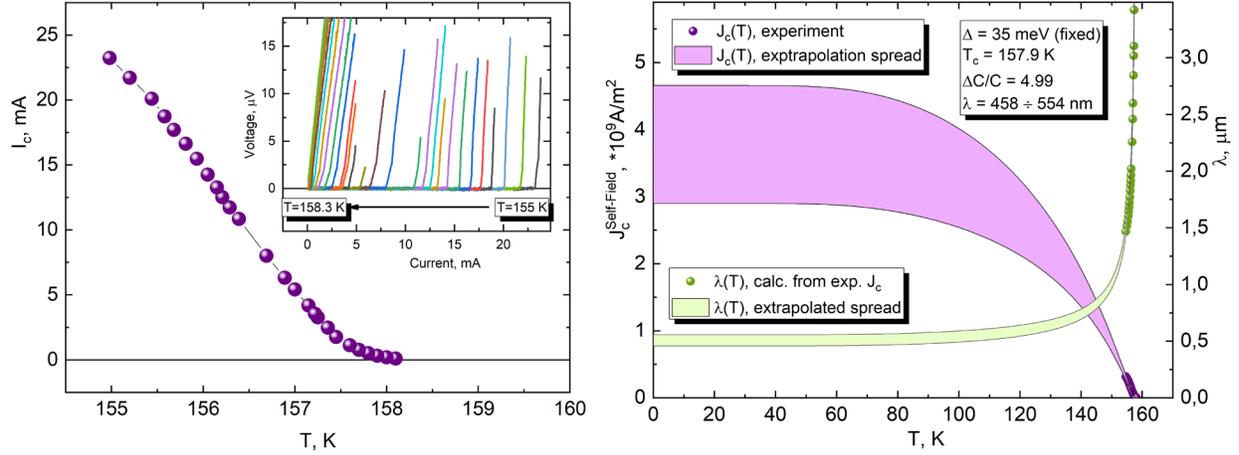

**Figure S3.** Left panel: self-field critical current temperature dependence $I_{c\_SF}(T)$. Inset: current-voltage characteristics for temperatures from 158.3 K to 155 K. Right panel: experimental (purple symbols) and theoretical (lines) dependences for critical current density $J_{C\_SF}(T)$ with the assumption of 2×4 µm and 2×5 µm cross sections. The green symbols and lines represent calculated values for $\lambda_L(T)$ using formulas (S1) for experimental and theoretical $J_{C\_SF}(T)$ data.